\documentclass[12pt]{article}

\usepackage{amsmath,amssymb,amsfonts,latexsym}
\usepackage{rotating}

\begin{document}
{ \begin{center}
\Large\bf {Qualitative analysis of certain generalized classes of quadratic oscillator systems}
\end{center}

\begin{center}
Bijan Bagchi\footnote{bbagchi123@gmail.com}, Samiran
Ghosh\footnote{sran$_{-}$g@yahoo.com}, Barnali
Pal\footnote{barrna.roo@gmail.com}, Swarup
Poria\footnote{swarupporia@gmail.com}
\end{center}

 \begin{center}
 { Department of Applied Mathematics}\\
{ University of Calcutta}\\
{ 92 Acharya Prafulla Chandra Road, Kolkata, India-700009}\\
 \end{center}
 \date{}

 \begin{abstract}
We carry out a systematic qualitative analysis of the two quadratic schemes of generalized oscillators recently proposed by C. Quesne [J.Math.Phys.\textbf{56},012903 (2015)]. By performing a local analysis of the governing potentials we demonstrate that while the first potential admits a pair of equilibrium points one of which is typically a center for both signs of the coupling strength $\lambda$, the other points to a centre for $\lambda < 0$ but a saddle $\lambda > 0$. On the other hand,  the second potential reveals only a center for both the signs of $\lambda$ from a linear stability analysis. We carry out our study by extending Quesne's scheme to include the effects of a linear dissipative term. An important outcome is that we run into a remarkable transition to chaos in the presence
of a periodic force term $f\cos \omega t$.
\end{abstract}

PACS:  {45.20.Jj, 45.50.Dd, 05.45.-a

\section{\label{sec:1}Introduction}
Nonlinear autonomous differential equations of second order have proved to be of much interest in the investigation of dynamical systems and nonlinear analysis \cite{GH, TAB, NM, WIG, LR, LC, AKT}. Consider a typical one belonging to quadratic Li\'{e}nard class as given by

\begin{equation}\label{1}
\ddot{x} + r(x)\dot{x}^{2} + s(x) =0
\end{equation}
where an overdot indicates a derivative with respect to the time variable $t$ and $r(x)$ and $s(x)$ are two continuously differentiable functions of the spatial coordinate $x$. The following specific forms of $r$ and $s$ which are odd functions of $x$, namely

\begin{equation}\label{2}
r(x) = -\frac {\lambda x}{1 + \lambda x^{2}}, \quad g(x) = \frac {\alpha^{2}x}{1 + \lambda x^{2}}, \quad \lambda > 0
\end{equation}
where $\alpha$ and $\lambda$ are nonzero real numbers, lead to a nonlinear equation defined by

\begin{equation}\label{3}
 (1 + \lambda x^{2})\ddot{x} - \lambda x \dot{x^{2}} + \alpha^{2}x = 0
\end{equation}
The Lagrangian relevant to (\ref{3})

\begin{equation}\label{4}
L = \frac {1}{2}\frac {1}{1 +\lambda x^{2}}(\dot{x^{2}} - \alpha^{2}x^{2})
\end{equation}
was studied by Mathews and Lakshmanan (ML) \cite{ML} long time ago in the search of a one-dimensional analogue of some quantum field theoretic model. One can observe that (\ref{4}) speaks of a $\lambda$-dependent deformation of the standard harmonic oscillator Lagrangian. Cari\~{n}ena et al \cite{Car1} also pointed out  that the kinetic term in  this Lagrangian is invariant under the tangent lift of the vector field $X_x(\lambda) = \sqrt{1+\lambda x^2} \frac{\partial}{\partial x}$.

The corresponding Hamiltonian represents a position-dependent effective mass system guided by the mass function of the specific type $m(x) = \frac {1}{1 + \lambda x^{2}}$. In the literature dynamics of several types of nonlinear systems have been found to be influenced by a position-dependent effective mass \cite{Car2, Ran, Cru, Bag, Mu,KS, GGH, Roy, SH}. From a physical point of view problems of position-dependent effective mass have relevance in describing the flow of electrons in problems of compositionally graded crystal, quantum dots and liquid crystals \cite{GKO,SL,BGHN}.

As is well known the nonlinear dynamics described by (\ref{3}) admits, in particular, a periodic solution for $x(t)$ given by the simple harmonic form

\begin{equation}\label{5}
  x(t) = A sin (\omega t +\phi)
\end{equation}
but with the restriction that the frequency $\omega$ is related to the amplitude $A$ by the constraint $\omega^{2} = \frac {\alpha^{2}}{1 + \lambda A^{2}}$. The amplitude thus depends on the frequency.

From the form of the Lagrangian (\ref{4}) it is easy to identify the corresponding potential $V(x)$ as given by

\begin{equation}\label{6}
  V(x) = \frac {1}{2} \frac {\alpha^{2}x^{2}}{1 + \lambda x^{2}}.
\end{equation}

Recently Quesne \cite{QS} extended the above potential to two
different types of generalizations by introducing a two-parameter
deformation by bringing in an additional phenomenological
parameter $\beta$ in the following manner:
\begin{equation}\label{7}
 (a)\quad  V_{I} = \frac {1}{2} \frac {\alpha^{2}x^{2} - 2\beta x}{1 + \lambda x^{2}}
\end{equation}

\begin{equation}\label{8}
 (b)\quad  V_{II} = \frac {1}{2} \frac {\alpha^{2}x^{2} - 2\beta x \sqrt {1 + \lambda x^{2}}}{1 + \lambda x^{2}}
\end{equation}
while keeping the kinetic part of (\ref{4}) unchanged. The respective Euler-Lagrange equations that follow from (\ref{7}) and (\ref{8}) read

\begin{equation}\label{9}
 (1 + \lambda x^{2})\ddot{x} - \lambda x \dot{x}^{2} + \alpha^{2}x - \beta {(1 - \lambda x^{2})} = 0
\end{equation}

\begin{equation}\label{10}
 (1 + \lambda x^{2})\ddot{x} - \lambda x \dot{x}^{2} + \alpha^{2}x - \beta \sqrt{ {(1 + \lambda x^{2})}} = 0
\end{equation}
Let us emphasize that the functional form corresponding to $s(x)$ in the above equations is not odd but of a mixed type consisting of a term that is odd and a term that is even. With an extra parameter $\beta$ at hand, quite expectedly, the potentials $V_{I}$  and $V_{II}$ allow dealing with richer behaviour patterns of the solutions of the corresponding Euler-Lagrange equations than the ones provided by (4) for the ML case. One of the guiding factors in this regard is the sign of the deformation parameter $\lambda$ that corresponds to different asymptotic behaviour of the two potentials $V_I$ and $V_{II}$ and the locations of the minimum for the latter. The first integral of the Euler-Lagrange equation when suitably confronted with the integration constant and the underlying discriminant also give crucial restrictions on the domains of the energy of the system for a physically viable solution.

In this paper we interpret the generalized nonlinear oscillators that are guided by $V_{I}$  and $V_{II}$ as examples of dynamical systems and perform a qualitative analysis to determine some interesting local properties for them. We also take up the issue of chaos in the presence of a periodic force term $f \cos\omega t$ for both the systems by fine-tuning some of the coupling parameters.
\section*{2\quad Local analysis of the potential $V_{I}(x)$}
Let us rewrite the Euler-Lagrange equation in the presence of $V_{I}$ as the following system of coupled equations
\begin{eqnarray}\label{11}
  \dot{x} & = & y \nonumber \\
\dot{y} & = &  \frac {\lambda x y^{2}}{1 + \lambda x^{2}} \quad - \frac {\alpha^{2}x}{1 + \lambda x^{2}} \quad + \frac {\beta(1- \lambda x^{2})}{1 + \lambda x^{2}}.
\end{eqnarray}
The resulting fixed points are readily identified to be located at the points  $(x_{1}^{*}, 0)$  and $(x_{2}^{*}, 0)$ where

\begin{equation}\label{12}
  x_{1,2}^{*}=(-\alpha^2 \pm \sqrt{\alpha^4+4\lambda \beta^2})/2\lambda\beta.
\end{equation}
The positivity of the discriminant leads to the restriction $\lambda>-\alpha^{4}/4\beta^{2}$.

Evaluating now the Jacobian matrix we obtain at the fixed points
\begin{equation}\label{13}
  J|_{(x_{1,2};0)}=\left(
                     \begin{array}{cc}
                       0 & 1 \\
                       A_{21} & 0 \\
                     \end{array}
                   \right)
\end{equation}
where
\begin{equation}\label{14}
A_{21}=(-\alpha^2+\lambda \alpha^2 x_c^2-4\lambda\beta x_c)/(1+\lambda x_c^2)^2
\end{equation}
and $ x_c=x_{1,2}^*$.

For $\lambda>0$, the respective eigenvalues of $J$ at $x_1^*$ and $x_2^*$ are
\begin{eqnarray}\label{15}
\pm i \sqrt{\frac{ 2\lambda \beta^2}{\sqrt{D}- \alpha^2}},\;\;
 \pm \sqrt{\frac{ 2\lambda \beta^2}{\sqrt{D}+ \alpha^2}}
\end{eqnarray}
where $D=\alpha^4+4\beta^2 \lambda >0$ implying  $\sqrt{D}>\alpha^2$. We therefore conclude that the point $(x_1^*,0)$
is a center while the other one, namely $(x_2^*,0)$, is a saddle point. The phase diagram of the system (\ref{11}) is plotted in Figure \ref{fg1}a
taking the parameter values $\lambda=0.5$ , $~\beta=0.34~$  $~\alpha=1.0~$.

In the case of $\lambda<0$ the two eigenvalues of the Jacobian
matrix (\ref{13}) at $x_1^*$ and $x_2^*$  are given by
\begin{eqnarray}\label{15a}
 \pm i\sqrt{\frac{ -2\lambda \beta^2}{\alpha^2 - \sqrt{D}}} ,\;\;
\pm i\sqrt{\frac{ -2\lambda \beta^2}{\alpha^2 + \sqrt{D}}}
\end{eqnarray}
where $D=\alpha^4+4\beta^2 \lambda >0$. $~\alpha^2>\sqrt{D}~$ as $ -\alpha^4/4\beta^2 \leq \lambda <0$.

Here too the equilibrium points $(x_{1}^{*},0)$ and $(x_{2}^{*},0)$ correspond to centers since  the eigenvalues of the Jacobian matrix
are  purely imaginary conjugate pairs for $\lambda<0$.  Figure \ref{fg1}b represents the phase portrait of the system (\ref{11}) for
 $\lambda=-0.5$ , $~\beta=0.34~$  $~\alpha=1.0~$. The Figure \ref{fg1}a and Figure \ref{fg1}b are
consistent.
\section*{3\quad Local analysis of the potential $V_{II}(x)$}
The dynamical system for the potential  $V_{II}(x)$ is  described by the  following set of equations

  \begin{eqnarray}\label{16}
 \dot{x} &=& y \nonumber \\
  \dot{y} &=& \frac {\lambda x y^{2}}{1 + \lambda x^{2}} \quad - \frac {\alpha^{2}x}{1 + \lambda x^{2}} \quad + \frac {\beta\sqrt{1+ \lambda x^{2}}}{1 + \lambda x^{2}}.
  \end{eqnarray}
The two equilibrium points correspond to  $(x_{1}^{*},0)$ and $(x_{2}^{*},0)$ where
\begin{equation}\label{17}
  x_{1,2}^{*}=\pm\frac{\beta}{\sqrt{\alpha^{4}-\lambda\beta^{2}}}
\end{equation}
subject to the positivity condition $\alpha^{4}-\lambda\beta^{2}>0$ i.e. $\lambda<\alpha^{4}/\beta^{2}$.
At the equilibrium points the Jacobian matrix takes the form
\begin{equation}\label{18}
  J|_{(x_{1,2};0)}=\left(
                     \begin{array}{cc}
                       0 & 1 \\
                       A_{21} & 0 \\
                     \end{array}
                   \right)
\end{equation}
where $A_{21}$ is given by
\begin{equation}\label{19}
  A_{21}=\frac{\alpha^{2}(\lambda x_{c}^{2}-1)-\lambda\beta x_{c}\sqrt{1+\lambda x_{c}^{2}}}{(1+\lambda x_{c}^{2})^2}
\end{equation}
and $x_{c}$ stands for  $x_{c}=x_{1,2}^{*}$ .

The eigenvalues of $J$ represent a pair of degenerate conjugate complex quantities namely,
\begin{equation}\label{20}
 \pm\frac{i(\alpha^{4}-\lambda\beta^{2})}{\alpha^{3}},~\pm\frac{i(\alpha^{4}-\lambda\beta^{2})}{\alpha^{3}}
\end{equation}
While the linear stability analysis indicates the nonhyperbolic
nature of the fixed points, as is well known such tests are not
always in conformity with the nonlinear analysis. Indeed, the
numerical simulation of the nonlinear system of (\ref{16}), as
shown in Figure \ref{fg3}, confirms that the linear stability
results correctly predict the behavior of one of the equilibrium
points namely $x_1^*$ but it fails in the case of the other
equilibrium point, $x_2^*$.
\section*{4\quad Inclusion of an external periodic forcing term}
$~~$ It is of  considerable interest to study the dynamics of
systems  (\ref{11}) and (\ref{16}), under the influence of the
external periodic forcing in the presence of additional damping,
so that the respective equation of motion become

 \begin{equation}\label{23}
 (1 + \lambda x^{2})\ddot{x} - \lambda x \dot{x}^{2} + \alpha^{2}x - \beta {(1 - \lambda x^{2})} + \gamma \dot{x} = f cos (\omega t)
\end{equation}

\begin{equation}\label{24}
{\mbox{and}~~} (1 + \lambda x^{2})\ddot{x} - \lambda x \dot{x}^{2} + \alpha^{2}x - \beta \sqrt{ {(1 + \lambda x^{2})}} +\gamma \dot{x} = f cos (\omega t).
\end{equation}
 The above nonautonomous systems can be rewritten as the following three dimensional autonomous nonlinear dynamical systems respectively

\begin{eqnarray}\label{25}
  \dot{x} & = & y \nonumber \\
\dot{y} & = &  \frac {\lambda x y^{2}-\gamma y}{1 + \lambda x^{2}} \quad - \frac
{\alpha^{2}x}{1 + \lambda x^{2}} \quad + \frac {\beta(1- \lambda x^{2})}{1 + \lambda x^{2}}+\frac{f\cos z}{1+ \lambda x^2} \\
\dot{z} & = & \omega \nonumber
\end{eqnarray}
and
\begin{eqnarray}\label{26}
  \dot{x} & = & y \nonumber \\
\dot{y} & = &  \frac {\lambda x y^{2}-\gamma y}{1 + \lambda x^{2}} \quad - \frac {\alpha^{2}x}{1 + \lambda x^{2}}
\quad + \frac {\sqrt{\beta(1+ \lambda x^{2})}}{1 + \lambda x^{2}}+\frac{f\cos z}{1+ \lambda x^2} \\
\dot{z} & = & \omega \nonumber
\end{eqnarray}
We have done numerical simulations of the system (\ref{25}) only.
The simulation results of the syatem (\ref{26}) are qualitatively
similar and hence not discussed.
\section*{5\quad Results and discussion}
\subsection*{5.1\quad Case-I:  No Dissipation ($ \gamma=0 $)}
We start with the dynamics of the ML-model in the absence of
dissipation. Thus we set $~\gamma=0~$ and consider a set of sample
parameter values $~\lambda=-0.5~$, $~\alpha=2.0~$, $~\gamma=0.0~$,
$~\omega=1.0~$, $~f=5.0~$ to plot the phase portrait in the
$xy$-plane. It reveals the quasi-periodic nature of the system
whose character survives for a wide range of inputs for the
external periodic forcing term. In Figure \ref{fg4} we have
plotted the phase diagram in the $xy$ plane of the system
(\ref{25}) for the parameter values $~\lambda=-0.5~$,
$~\alpha=2.0~$, $~\gamma=0.0~$, $~\omega=1.0~$, $~f=5.0~$
corresponding to  the test values of $~\beta~$ as given by
$~\beta=0.001~$, $~\beta=0.01~$ and $~\beta=0.1~$. These values of
$~\beta~$ correspond to small and moderate departures from the
ML-model. Figure \ref{fg4}a illustrates the case of
$~\beta=0.001~$  while the Poincar\'{e} first return map of the
same is plotted in Figure \ref{fg4}b. We observe that the first
return map data lie on 3 smooth closed curves confirming the
existence of the quasiperiodic behaviour of the system. The latter
behavior persists for the choice of $~\beta=0.01~$ also.  The
phase diagram for $~\beta=0.1~$ keeping the other parameters fixed
is exhibited in Figure \ref{fg4}c. However, the first return map
as presented in Figure \ref{fg4}d shows a drastic change. The
irregular scattering of points in the latter figure points to the
existence of chaos in the system. We therefore observe the
sensitivity of the system (\ref{25}) to the choice of the
$~\beta~$-values as evidenced by the transition from
quasi-periodicity to chaos as the value of  $~\beta~$ is changed
from $~\beta=0.001~$ by two orders of magnitude.

Next we discuss for the system (\ref{25}) the phase diagrams and
those for Poincar\'{e} return maps for different values of forcing
amplitude $~f~$ for the data set $~\lambda=-0.5~$, $~\alpha=2.0~$,
$~\beta=0.1~$, $~\gamma=0.0~$, $~\omega=1.0~$. These are
summarized in Figure \ref{fg5}. In Figure \ref{fg5}a we  have
plotted the phase portrait for $~f=0.0~$ and the corresponding
Poincar\'{e} first return map is shown in Figure \ref{fg5}b.  The
existence of only one point in the Figure \ref{fg5}b proves the
existence of periodic orbit in case of no external forcing. In
Figure \ref{fg5}c the phase diagram  of the system is presented
for $~f=3.0~$ and the corresponding Poincar\'{e} first return map
is given in Figure \ref{fg5}d. The return map data lie on smooth
closed curves and hence confirms the existence of quasiperiodic
behaviour of the system for $f=3.0$. We have plotted the phase
portrait for $~f=5.0~$ in Figure \ref{fg5}e and the corresponding
return map data are plotted in Figure \ref{fg5}f. The irregular
set of points in the whole  plane guarantee the existence of chaos
in the system for $~f=5.0$. Therefore  as in the previous case the
increase of forcing amplitude also produces chaotic motion via the
quasiperiodic route.
\subsection*{5.2 \quad Case-II: Dissipative Case ($ \gamma \neq 0 $)}
We now turn to the inclusion of the dissipation effects in the
model (\ref{25}). Towards this end we keep the following fixed
sample values of  $~\lambda=-0.5~$, $~\alpha=2.0~$, $~\beta=0.1~$,
$~\omega=1.0~$ and $~f=5.0~$ and vary the dissipation parameter
$\gamma$. We consider an assorted set of values for $\gamma$
ranging from very small to moderately large values. We represent
in Figure \ref{fg11}a the phase portrait in the xy plane for
$~\gamma=0.002~$ and the corresponding Poincar\'e map in the
Figure \ref{fg11}b. We find a set of randomly distributed points
that tells us about the chaotic character inherent in the system
(\ref{25}). Next for $~\gamma=0.02~$, from   the phase portrait of
system (\ref{25}) plotted in the Figure \ref{fg11}c and
corresponding Poincar\'{e} map in the Figure \ref{fg11}d, we
notice a finite number of closed curves that confirm the existence
of quasiperiodic oscillations. Figures \ref{fg11}e and \ref{fg11}f
display the phase portrait and time evolution of the variable $y$
for an order of magnitude higher value of $~\gamma=0.1~$. Time
evolution of the corresponding phase portrait of Figure
\ref{fg11}f guarantees the presence of a periodic behaviour for
$~\gamma=0.1~$. In Figure \ref{fg10} we provide the bifurcation
diagram of the variable $y$ with respect to the parameter
$~\gamma~$. It is clear that, with the increase of dissipation,
the system undergoes a  transition from chaos to quasiperiodicity
and then settles to a periodic behavior.
\section*{6\quad Summary}
In this paper we studied the dynamical behavior of two quadratic
schemes of generalized oscillators recently proposed by Quesne. We
performed a local analysis of the governing potentials to
demonstrate that while the first potential admits a typically
center-like equilibrium point  for both signs of the coupling
strength $\lambda$, the second potential only admits  to a centre
for $\lambda < 0$ but a saddle $\lambda > 0$. The second potential
however reveals, from a linear stability analysis,  only a center
for both the signs of $\lambda$. We have extended Quesne's scheme
to include the effects of a linear dissipative term and shown how
inclusion of an external periodic force term changes the
qualitative behavior of the underlying systems drastically leading
to the possible onset of chaos.

\begin{figure}
\includegraphics[width=0.4\textwidth]{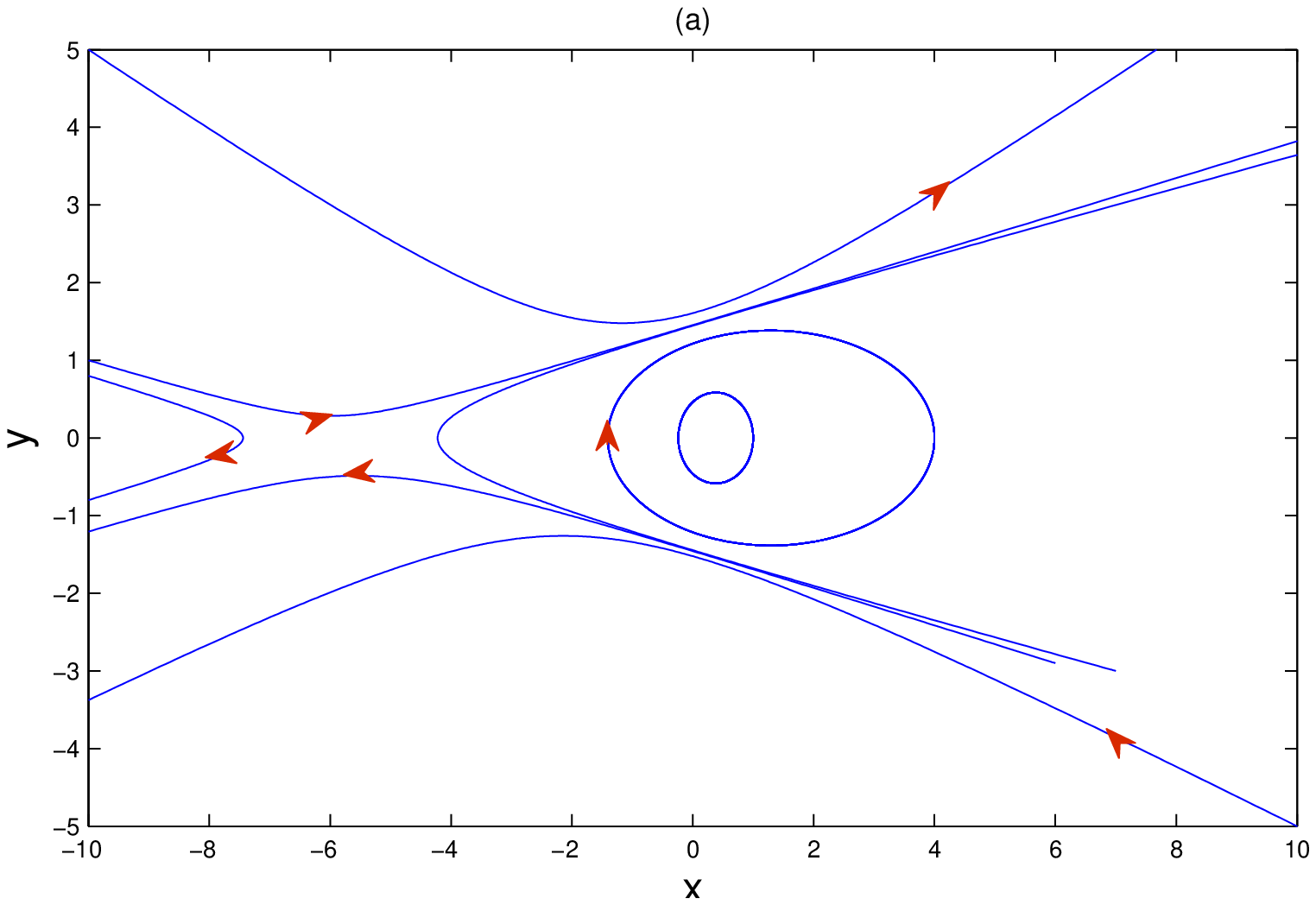}
\includegraphics[width=0.42\textwidth]{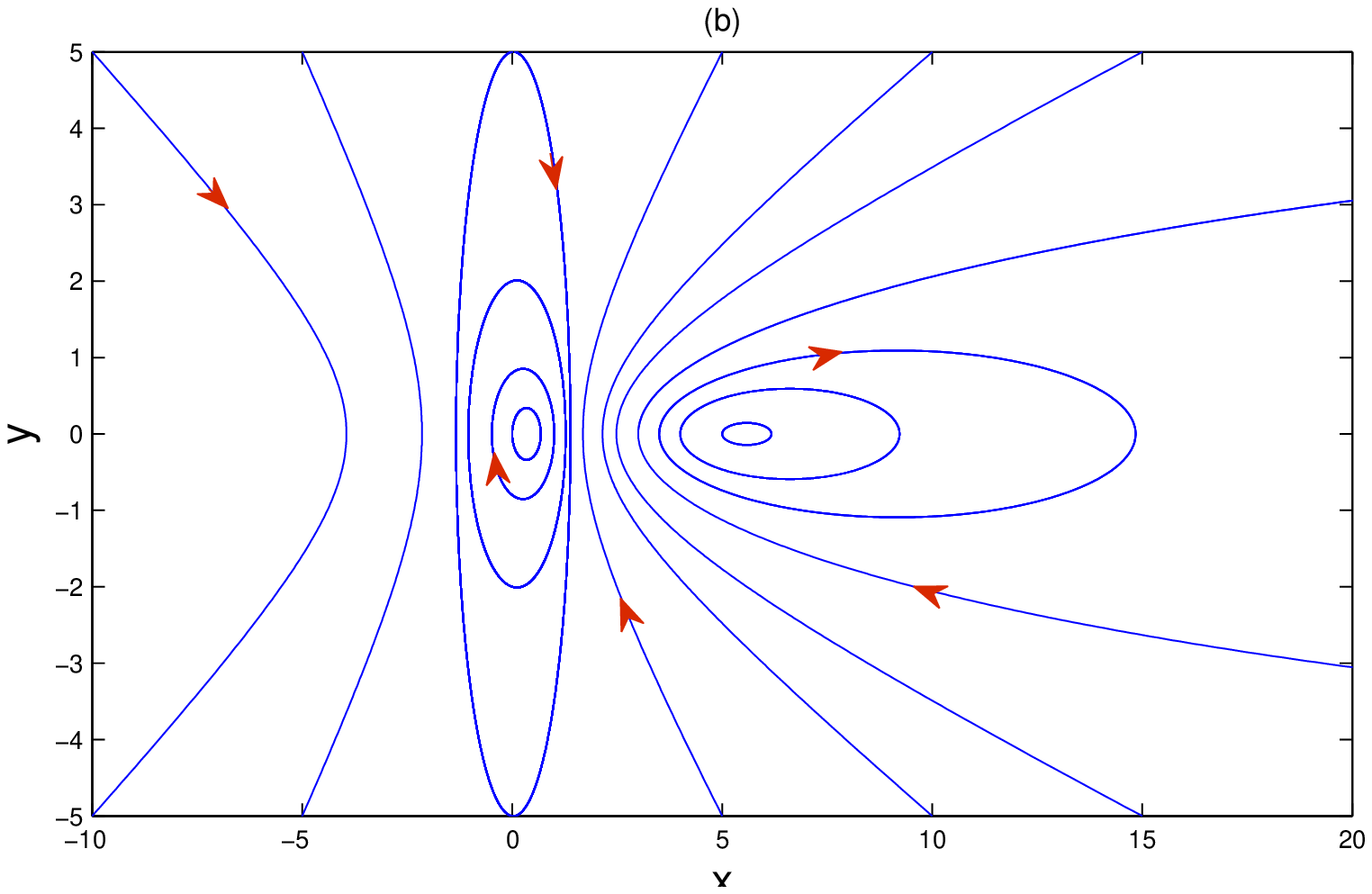}
\caption {Phase diagram of the system (\ref{11}) for (a)$~\lambda=0.5~$, (b)$~\lambda=-0.5~$  with $~\beta=0.34~$, $~\alpha=1.0~$.}
\label{fg1}
\end{figure}

\begin{figure}
\begin{center}
\includegraphics[width=1.2\textwidth]{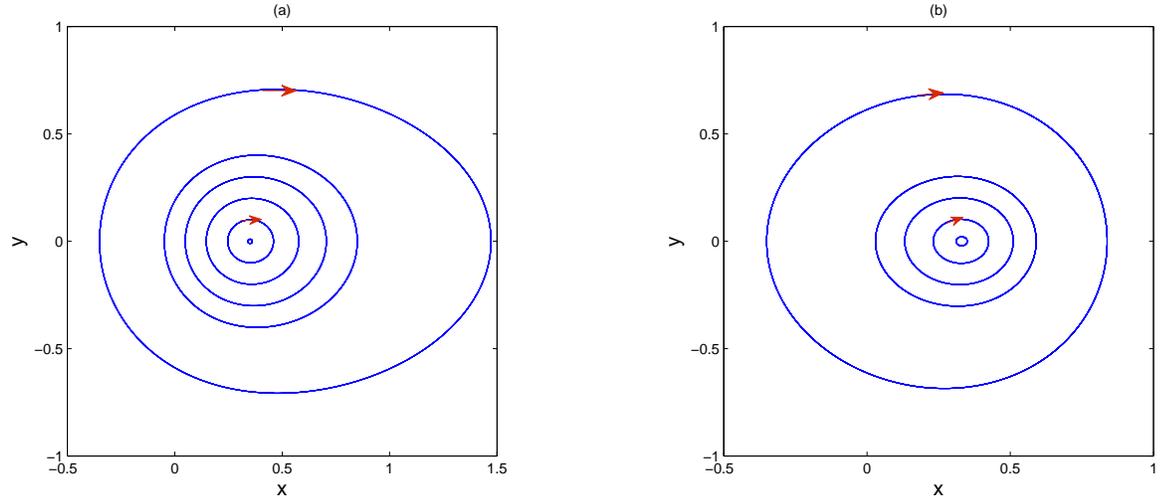}
\caption {Phase portrait of the  system (\ref{18}) for (a)$~\lambda=0.5~$ ,(b) $~\lambda=-0.5~$; with $~\beta=0.34~$, $~\alpha=1.0~$.}
\label{fg3}
\end{center}
\end{figure}

\begin{figure}
\begin{center}
\includegraphics[width=1.2\textwidth]{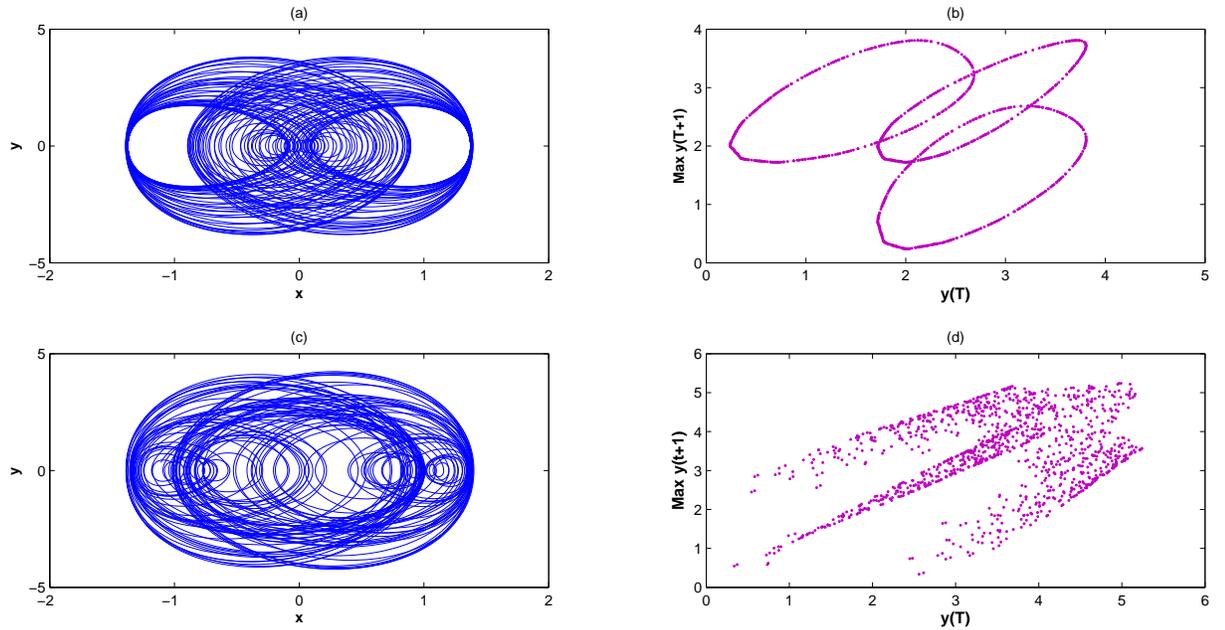}
\caption {Phase portrait of the system (\ref{25}) with $~\lambda=-0.5~$, $~\alpha=2.0~$, $~\gamma=0.0~$, $~\omega=1.0~$, f=5.0 for different value of $~\beta~$ (a) $~\beta=0.001~$, (c)$~\beta=0.1~$; (b) and (d) represent corresponding Poincare first return map.}
\label{fg4}
\end{center}
\end{figure}

\begin{figure}
\begin{center}
\includegraphics[width=1.2\textwidth]{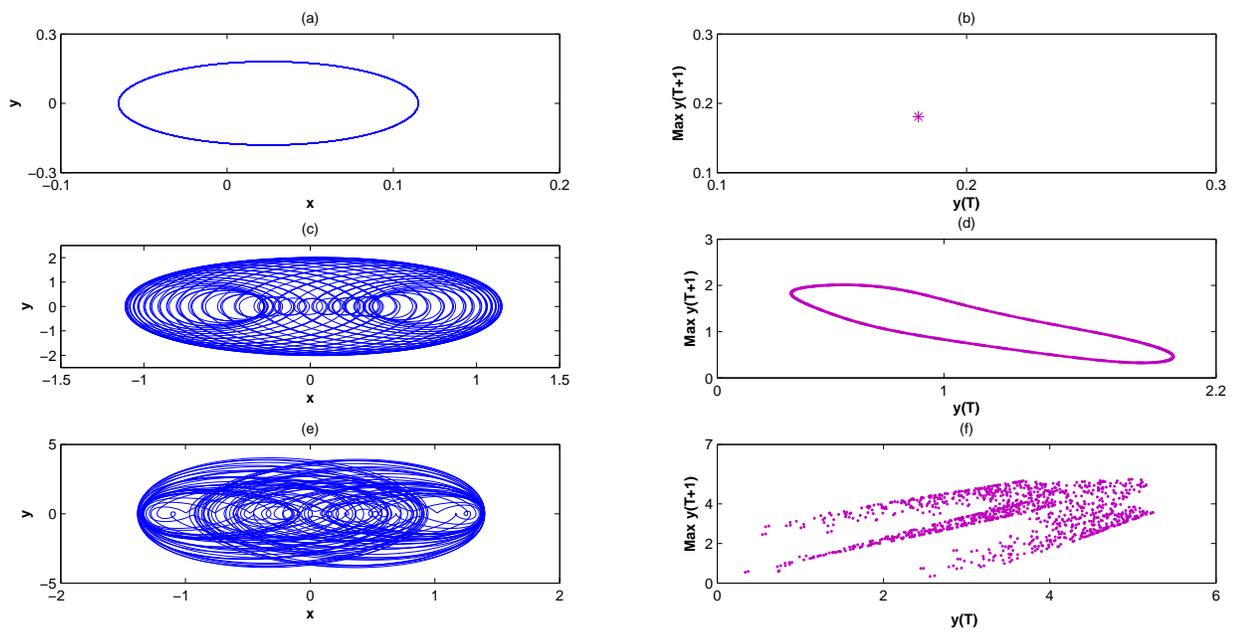}
\caption {Phase portrait of the system (\ref{25}) with $~\lambda=-0.5~$, $~\alpha=2.0~$, $~\beta=0.1~$, $~\gamma=0.0~$, $~\omega=1.0~$ for different value of $f$ (a) f=0.0, (c) f=3.0, (e) f=5.0 ; (b),(d) and (f) represent corresponding first return map.}
\label{fg5}
\end{center}
\end{figure}

\begin{figure}
\begin{center}
\includegraphics[width=1.2\textwidth]{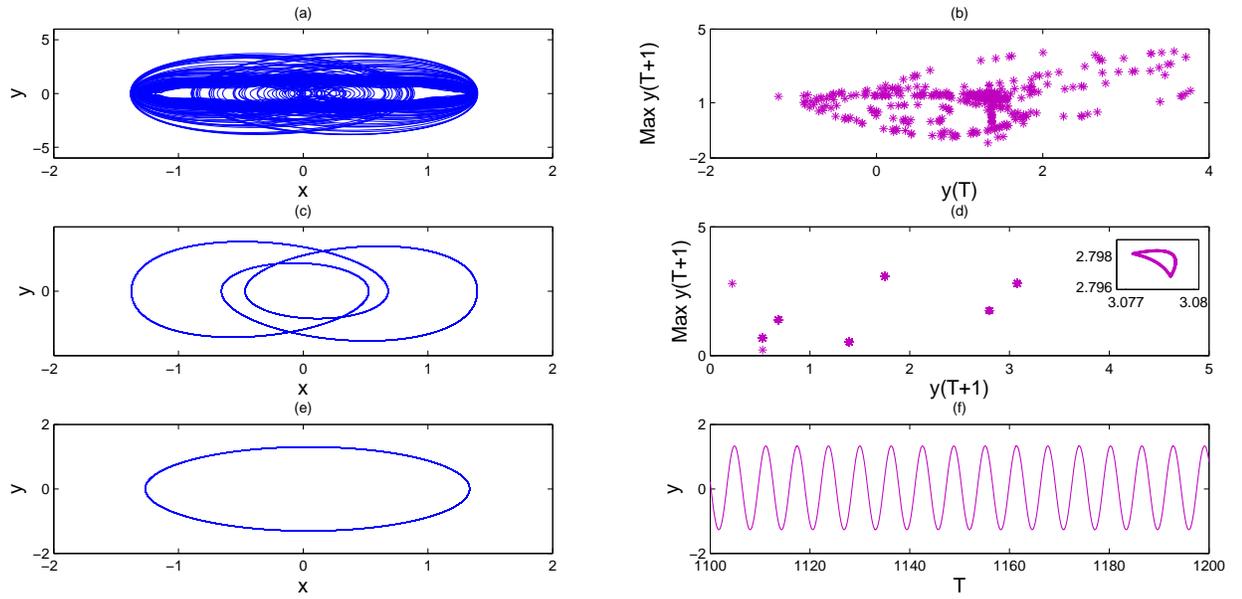}
\caption { Phase portrait of the system (\ref{25}) with $~\lambda=-0.5~$, $~\alpha=2.0~$, $~\beta=0.1~$, $~\omega=1.0~$ and $~f=5.0~$ for different value of $~\gamma~$ (a) $~\gamma=0.002~$,(c) $~\gamma=0.02~$, (e) $~\gamma=0.1~$; (b),(d) represent corresponding first return map; (f) corresponding time series}
\label{fg11}
\end{center}
\end{figure}

\begin{figure}
\begin{center}
\includegraphics[width=0.9\textwidth]{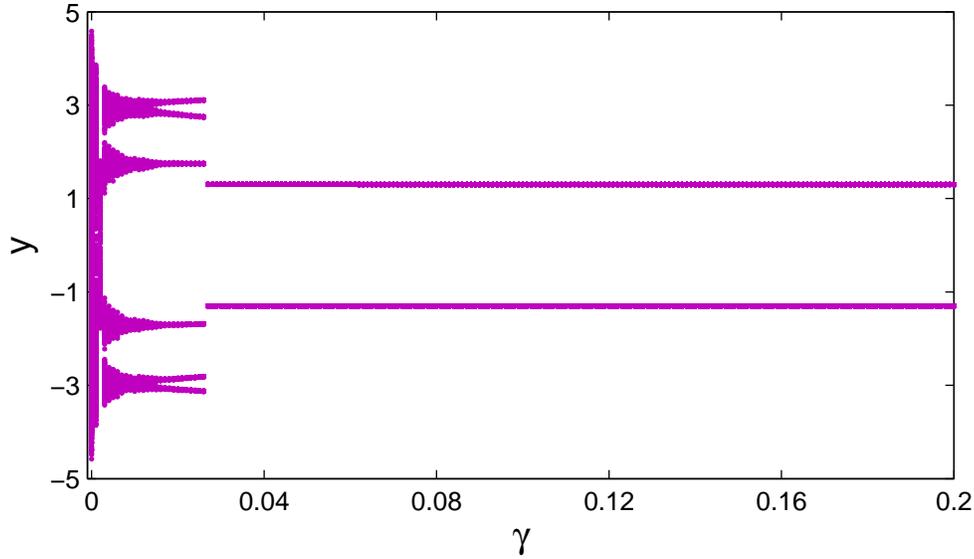}
\caption { Bifurcation diagram of y with respect to parameter $~\gamma~$  of the system (\ref{25}) with $~\lambda=-0.5~$, $~\alpha=2.0~$, $~\beta=0.1~$, $~\omega=1.0~$ and $~f=5.0~$.}
\label{fg10}
\end{center}
\end{figure}

\end{document}